\documentstyle[11pt,aaspp4]{article}
\newcommand{\aas}{{ A\&AS,}~}

\newcommand{\kms}{\ifmmode \hbox{ \rm km s}^{-1} \else{ km s$^{-1} $}\fi}
\newcommand{\kmsp}{~km~s$^{-1}$.\ }
\newcommand{\kmsc}{~km~s$^{-1}$,\ }

\newcommand{\app}{$\approx$~}

\newcommand{\etal}{et~al.\ }

\newcommand\simlt{\lesssim}     % In amstex, works in AASTEX
\newcommand\simgt{\gtrsim}      % In amstex, works in AASTEX
\newcommand{\hub}{ $h^{-1}_{75}$ }
\newcommand{\parcm}{.\hspace{-0.09cm}$'$}
\newcommand{\parcs}{.\hspace{-0.09cm}$''$}

\singlespace

\begin{document}

\slugcomment{To Be Submitted to the Astronomical Journal}
\lefthead{Pinkney et al.}
\righthead{Substructure in WAT Clusters I.}

\title{\bf SUBSTRUCTURE IN CLUSTERS CONTAINING WIDE-ANGLE TAILED
RADIO GALAXIES. I.  NEW REDSHIFTS}

\author{JASON PINKNEY\altaffilmark{1}}
\affil{ Department of Astronomy, University of Michigan,
500 Church St., Ann Arbor, MI 89109 \\
jpinkney@astro.lsa.umich.edu}

\author{JACK O. BURNS\altaffilmark{1}}
\affil{Office of Research and Department of Physics and Astronomy,
University of Missouri, Columbia, MO 65211 \\
burnsj@missouri.edu }

\author{MICHAEL J. LEDLOW\altaffilmark{1}}
\affil{ Department of Physics and Astronomy, University of New Mexico,
Albuquerque, NM 87131 \\ 
mledlow@gemini.edu }

\author{PERCY L. G\'{O}MEZ\altaffilmark{1}}
\affil{ Department of Physics and Astronomy, Rutgers University 
PO Box 849, Piscataway, NJ 08855 \\
percy@physics.rutgers.edu }
\and
\author{JOHN M. HILL} 
\affil{Steward Observatory, University of Arizona, Tucson, AZ 
85721-0065 \\
jhill@as.arizona.edu }
\altaffiltext{1}{Former address: Department of Astronomy,
New Mexico State University, Box 30001/Dept. 4500, Las Cruces, NM 88003}
 
\begin{abstract}
We present new redshifts and positions for 635 galaxies in nine
rich clusters containing Wide-Angle Tailed (WAT) radio galaxies.  
Combined with existing data, we now have a sample of 18 WAT-containing
clusters with more than 10 redshifts.
This sample contains a substantial portion of the WAT clusters in the 
VLA 20 cm survey of Abell clusters, including 
75\% of WAT clusters in the complete survey (z$\leq$0.09), 
and 20\% of WAT clusters with z$>$0.09. 
It is a representative sample which should not contain biases other 
than selection by radio morphology.
We graphically present the new data using histograms and sky maps.
A semi-automated procedure is used to search for emission lines 
in the spectra in order to add and verify galaxy redshifts. We find
that the average {\em apparent} fraction of emission line galaxies is
about 9\% in both the clusters and the field.
We investigate the magnitude completeness of our redshift surveys
with CCD data for a test case, Abell 690.  This case indicates
that our galaxy target lists are deeper than the
detection limit of a typical MX exposure,
and they are 82\% complete down to R=19.0.
The importance of the uniformity of the placement of fibers on
targets is posited, and we evaluate this in 
our datasets.  We find some cases of non-uniformities which may
influence dynamical analyses.
A second paper will use this database to look for correlations between
the WAT radio morphology and the cluster's dynamical state.
\end{abstract}
 

\section{Introduction}

Wide-angle tailed radio galaxies (WATs) are a powerful type of FR-I
(``edge-darkened", \cite{FR74}) radio galaxy
associated with first-ranked ellipticals in clusters. 
Generally, the tails of a WAT bend in a common direction, giving 
the source an overall {\sc V}, or {\sc C} shape (see   % \cite{OOE} 
O'Donoghue \etal 1990 for a radio survey).  
The mechanism which bends the WATs is poorly understood.
Ram pressure caused by motion through the intracluster medium (ICM) 
was suspected from the time of discovery (\cite{OR76}).
The sticking point in this hypothesis is that 
\app 1000 \kms\ relative velocities may be required for bending
(\cite{Eil84}), and the host galaxies, typically gE, D 
and cD, should have much smaller peculiar velocities with respect to 
their cluster (\cite{Mal92}, \cite{Bu86}).

A subcluster merger hypothesis has emerged linking the bent
morphology of WATs to the dynamical state of the host 
cluster (Pinkney \etal 1993, hereafter \cite{Pi93}, 
Gomez et al 1997,
\cite{Rot96}, \cite{Lok95}).
In it, the WAT acts as a stationary weathervane, its tails indicating
the local flow of ICM in the stormy environment of a
cluster-subcluster merger (\cite{Burns98}). If this hypothesis is true,
then WAT's should be beacons of cluster mergers.

This hypothesis is difficult to confirm for
a {\em single} cluster: only {\em radial} peculiar velocities can
be ruled out for the WAT galaxy, and the optical 
signatures of merger can be subtle during post-merger 
epochs when the ICM motion is still sufficient to shape the WAT
(\cite{Lok95}; \cite{Pi96}). 
A large database can provide
statistical leverage to test the subcluster-merger hypothesis.
In particular, a distribution of WAT radial peculiar velocities can 
constrain
the true WAT velocities, thereby testing the moving-WAT hypothesis. 
Moreover, the occurrence rate of significant
substructure can be compared to a control sample of radio-quiet clusters.
In this paper, we present new velocities 
and positions of over 600 galaxies in 9 WAT clusters.  
This will be combined with published data for 9 additional WAT clusters.
These data will be used to probe the connection between
WAT bending and cluster evolution in paper II.
In this paper, section 2 will contain the sample selection criteria.
Section 3 describes the observations and reductions. 
In section 4, we compare our velocities to published ones and
discuss the emission line content. Finally, in section 5 we
examine the completeness and spatial uniformity of our
redshift surveys.

\section{WAT Cluster Sample Selection}

Our primary selection criterion for clusters was radio morphology.
Clusters with high resolution radio data were preferred; if
there is a morphological trait among WATs that correlates
with their dynamical state, we want it to be apparent in
the existing maps.
We also required galaxies with m$_{V} \lesssim 18.0$ 
(Pinkney et al. 1994, hereafter Pi94),
and $ \delta < 72\arcdeg\ $ for observation with the Steward Observatory
2.3-m telescope + MX spectrometer.

Our resources for sample selection include the VLA survey of
11 WATs by O'Donoghue \etal (1990, hereafter OOE), and the
20 cm VLA Survey of Abell Clusters
(\cite{OL97} and references therein).
This survey contains 38 radio sources informally
classified as ``WAT" or ``WAT?", 8 of which comprise 2.3\% of the 
complete cluster sample (0.0 $\leq$ z $\leq$ 0.09).
The survey uses the VLA C-array (1.4 GHz), with adequate
resolution (3-15$''$) to reveal the WAT morphology out to z$\sim$0.2.
The VLA maps in \cite{OOE} generally have higher resolution and sensitivity.
Therefore, we attempted to draw as many clusters from this survey
as possible.  However, several of the OOE clusters are beyond the redshift
and declination limits of the MX spectrometer.
We emphasize that we did not select WATs by the apparent angle
formed by their tails as it might bias the peculiar velocity
measurements.
Fortunately, the OOE survey included a wide variety of 
morphologies, so that WATs with strongly bent and weakly
bent tails were present.  
 
We obtained new optical spectra for 7 Abell clusters from
OOE: A98, A160, A690, A1446, A1684, A2214, and A2462, and for
1 Abell cluster not in OOE, A2220.
Together with the published databases for A115 (\cite{BHG83}), 
A400 (\cite{A400}),
A623 and A2304 (\cite{Gom98}),       % Thesis
A1346 (\cite{Sling98}),
A1940 (\cite{Huc92}),             % CfA
A1569 (Gomez et al 1997),
and A2634 (\cite{Pi93}, \cite{Sco95}), 
we have redshift databases for 16 Abell clusters with WATs.

In addition, we observed a {\it poor cluster }
containing the WAT 1313+073 (\cite{Pat86}). We will also include
the published database for the poor cluster containing the
WAT 1919+479 (\cite{Pi94}).  Both are good examples
of the WAT morphology with excellent radio maps available.
We argue that inclusion of these poor clusters is appropriate
in this study because
the X-ray and radio properties of poor and rich clusters overlap 
(\cite{Pr91}, \cite{Burns96}), and redshift-corrected membership counts of poor 
clusters overlap those of richness class 0 and 1 Abell clusters
(\cite{And94}).  

The radio properties of our combined sample of 18 WATs are shown in 
Table \ref{tabwatrad}.  
This sample contains 6/8 (75\%) of WAT clusters in the complete 
20 cm VLA survey (z $\leq$ 0.09), 
and 7/30 (23\%) of the WAT clusters in the high redshift portion
of the VLA survey.
The ROSAT X-ray data for five of these clusters were discussed in \cite{Gom97}. 
The quantities ``Extent", r/A$_C$, and log(P$_{1400}$) in Table \ref{tabwatrad} 
might all conceivably be correlated with global cluster properties.  We will 
return to these in paper II.

\placetable{tabwatrad}

% Section 3
\section{Multifiber Spectroscopy}

	The spectral observations of cluster galaxies were made
with the MX multifiber spectrometer (\cite{HL86})
on the Steward Observatory 2.3-meter telescope.
The MX uses mechanical probes to position
2$''$ diameter fibers in the focal plane with 0\parcs5 precision.  
The useful field diameter is 45$'$. Up to
32 fibers are assigned to galaxies, while 29 are placed on the sky.  

Our initial detector was a TI 800x800 CCD which became unreliable after our
Dec. 31, 1991 run (Table 2).  
For the remaining runs, we used Loral 1200$\times$800 CCDs.  
The Loral chips had a superior full well capacity, readout noise, and
Q.E. (which exceeded 90\% at 4000\AA ).  
Our first Loral chip had a problem with excessive `cosmic rays' 
caused by a radioactive coating on the chip.  On the next run, 
the second Loral chip lacked the radioactive coating.  
However, it had a problem with large gradients in bias which varied 
with time.
This problem was handled in reduction by 
replacing standard bias subtraction with subtraction of a 
surface fit to inter-aperture counts.
The old 300 lines/mm diffraction grating was switched with a
400 lines/mm grating (4890\AA\ blaze) for the Dec 1992 run.
This changed the dispersion from 3.6 to 2.6 \AA\ pix$^{-1}$ and the 
spectral resolution improved from 8.3 \AA\ to 6.1 \AA.
After two useful runs with the second chip, a third
Loral chip was used which had a stable bias level.
Some sample spectra from our surveys are shown in Figure \ref{fig1}.
The spectral range was nominally reduced to 3800-6900 \AA\ for 
cross-correlation.

\placetable{mxlog}

\placefigure{fig1}

	Our basic CCD reduction, spectral extraction and cross-correlation
procedures are much the same as in \cite{Pi93} and Pi94.
Post-correlation reduction includes a new, objective algorithm for
combining the velocities produced by 21 templates.
This algorithm uses the $r$ value of the CCF 
(cross-correlation function, \cite{TD79}),
and the number of templates agreeing on the same velocity
(within a 600\kms\ range) as criteria for including or excluding
velocities in an average.
We use three categories for a galaxy's final velocity measurement,
category 1 (hereafter C1), those with an average $r$ value $\geq 3.5$ and
at least 4 templates in agreement; category 2 (hereafter C2), those with
$3.0 \leq {r} < 3.5$  and at least 4 templates in agreement;
and ``invalid", those which do not fall into the above categories.
The C2 category includes velocities which are $\simgt$ 50\% likely
to be determined from the correct peak in the CCF. Therefore, 
C2 galaxies can be classed as cluster members or non-members 
with some accuracy, but should be omitted from analyses of the
velocity distribution.
The $r$ value was also used in combining redundant observations
of galaxies: spectra with low $r$ were excluded from the velocity
determination if high $r$ spectra existed.
The final velocities and positions are tabulated in Table \ref{bigtab}.
The fourth and fifth columns contain the C1 and C2 velocities,
respectively. 
Galaxies with invalid measurements, with no measurements,
and stars, are omitted.

To determine the velocity errors (column 6, Table \ref{bigtab}),
we fitted a curve to the relation between velocity deviation and the $r$
value of the CCF (similar to \cite{Pi94}, \cite{HO93}).
We used 251 repeat spectra of galaxies from 4 clusters 
taken with the Loral CCD systems for this calibration.  
The resulting fit is $\epsilon(r) = 486.7(1.0+r)^{-1}$.
When three or more good spectra
were obtained, the standard deviation was used as the velocity error.
If the calculated errors were less than 20 \kmsc
the final velocity error was set to 20 \kmsp
The last two rules also applied to emission line velocities (see \S 4.2).
An emission line redshift would only be adopted if the cross-correlation 
results had a larger scatter or $r \simlt 4.0$. 
If only one spectrum existed, the standard deviation of velocities 
derived from each emission line was used. All C2 velocities were
assigned an error of 200 \kmsc  to discourage their use
for purposes that require greater velocity precision.

\placetable{bigtab}

\section{The WAT Velocity Database}

\placefigure{fig2}

A collage of histograms is shown for our nine new datasets in
Figures \ref{fig2} and \ref{fig3}. It is clear that the WATs are associated
with real systems in redshift space.  
% Velocity substructure will be addressed in Paper II.
Also, visual inspection of these figures reveals obvious velocity
substructure in clusters such as A98, and A160.  In Table \ref{means}
we summarize the cluster velocity distributions.  Here we use the
traditional mean and standard deviation and apply them to the 
velocities within $\pm$4000 \kms\ of the WAT.  These estimators of
velocity centroid and dispersion are notoriously poor in the presence of
outliers or substructure (Beers \etal 1990).  In Paper II, we
will account for substructure in the datasets and apply more robust 
estimators to the velocity distribution.

\placefigure{fig3}

\subsection{Comparison with literature}

As a check on our velocity measurements, we have compared 
our results to those published for A2462 (also called A3897, 
\cite{Kat98}) and A98 
(\cite{BGH}; \cite{FD77}; \cite{Zab90}) in Table \ref{vcomp}.  
Redshifts for A98 were first obtained by \cite{FD77} (FD) and
were added to by \cite{BGH} (BGH). 
The average difference between our velocities and those of 
BGH and FD combined is -67\kmsc with $\sigma$=113\kms (N=7).  
This standard deviation is consistent with random errors in velocity of 80\kmsp
Zabludoff \etal (1990) later took the data added by
BGH and re-reduced it with a new template.  
The difference between our velocities and the velocities in
Zabludoff \etal are 40\kms, with $\sigma$=157 (N=5). 
Two velocities were measured by \cite{FD77} and ourselves,
but not BGH, they differed by -193 and -132\kmsp

We wish to add to our dataset the 21 BGH/FD galaxies which we did not observe.
The measured zero-point shift (-67 \kms) 
is only marginally inconsistent with no shift (1.6$\sigma$), so we will
not attempt to correct for it.
We have precisely re-measured the positions of the BGH galaxies 
on our quick-V frame based on their finding chart. We include them 
in Table \ref{bigtab}, and
give them ID's that begin with ``0".

Table \ref{vcomp} compares the 12 velocities in Abell 2462 which were
measured by both ENACS (ESO Nearby Abell Cluster Survey) and us.
% Their precise positions allowed easier cross-identification.
The mean difference (after removing the 325\kms\ CMB correction used
by ENACS) was -98.8\kms\ with a standard deviation of 111.2\kmsp
% Although our errors are sometimes larger than those of ENACS, our
% TDR values were also larger, so we retained our own measurements.
One velocity, ENACS \# 12, was adopted since we had no measurement.
The ENACS galaxies are labeled in the notes column of Table \ref{bigtab}.
Our velocities are in reasonable agreement with published values.

\subsection{Emission Line Galaxy Survey}

We searched our data for emission line galaxies (ELGs). 
The primary goals were to strengthen the accuracy of the velocity 
measurements and to increase cluster membership.  
The codes in column 7 of Table \ref{bigtab} indicate
the presence of emission lines in the spectra.  

Emission line redshifts were measured in a two-stage process.
The first stage identified potential redshift systems
using IRAF-based scripts and a Fortran program. 
Each spectrum was continuum-subtracted and its positive features
over \app$2\times$RMS were identified. Each feature was fitted with a
Gaussian. The features were then searched for redshift systems containing
any 2 or more of 15 lines produced by [OI], [OII], [NeIII], [OIII], 
[NI], [NII], [SII] or the Balmer series.
To reduce spurious results, the program rejected features that were:
1) too close to 4 major night sky lines, 2) too thin (FWHM$\leq 3.8$ \AA),
3) too wide (FWHM$\geq 25$ \AA), or 4) too low in flux ($<$ 20 DN).  
The program ranked the systems based
upon how well their features agree in redshift, and the number of features
they contain.  

In the second stage, the output from the automated search was used as a 
guide to judge the final validity of each system.  We judged redshift systems
based on the line id's, fluxes, FWHM, redshift agreement 
with each other and agreement with cross-correlation results.
We also visually inspected the candidates to look for lines missed
by the automated search, and to choose between the redshift systems.
A spectrum had to have at least 2 lines of [OII], [OIII] $\lambda$5007, 
H$\beta$, or H$\alpha$ to be acceptable, while lines of [OI], 
[OIII] $\lambda$4363, [NeIII], and [NI] were insufficient.
Moreover, we expected line ratios to be typical of astrophysical
environments, e.g., the [OIII] $\lambda$4959 
line flux should not be much larger than that of $\lambda$5007, and H$\beta$
should not be larger than H$\alpha$.  Convincing lines typically had 
FWHM $\simgt$ 6 \AA\ and rarely exceeded 15 \AA .  The sensitivity
limit in equivalent width was roughly -5 to -10\AA .
Real line systems typically had internal scatter $\simlt$ 100\kmsp
An example of a convincing ELG spectrum is shown in Fig 1c.

\placetable{elg}

We show the results of our ELG survey in Table \ref{elg}.
Here we see a range of 0 - 20\% in the percentage of cluster
members with emission, with an average of 8.6\%. As a comparison, 
Biviano \etal (1997) find \app 16\% in the ENACS database,
although they have perhaps less stringent criteria (1 emission line
is sufficient).  Each of our percentages should be
considered a lower limit to the true ELG percentage because many 
factors reduce our detection of emission lines.  
The positioning of the H$\alpha$ line within the spectral
range was the main factor (see Table \ref{elg}).  
Abell 160 has the highest ELG\% and this is probably related to the 
frequent inclusion of H$\alpha$ within the spectral range.
Another factor is the confusion of galaxy emission with 
night sky lines.  Some members of A98 and A1446 will have
their [OIII] redshifted near $\lambda$5577, and the automated
program will reject any line within a window centered on that line.
Despite this, the ELG\% for A1446 appears high. However, the 
ELG fraction for non-members in the A1446 field is also high, suggesting 
that excellent data quality is the cause rather than
cluster properties.
In general, our data are not sufficiently homogeneous
to make direct comparisons between ELG fractions in our clusters.
The bottom row of Table \ref{elg} indicates
that the apparent fraction of ELGs is roughly the same among our member
and non-member (``field") samples.  This coincidence results
from our particular detection limits for ELG's and non-ELG's.
\cite{Biv97} find an {\it apparent}
fraction of ELGs among the field about $2\times$ greater than 
inside clusters (a real field-cluster difference is expected
because of morphological segregation). 
Given that their search criteria are less stringent than ours, 
their sensitivity to ELG's is higher in both environments.
But the {\em fraction} of ELGs will be disproportionately higher in the field
because it contains mostly spirals. So it is not surprising that
our results differ.

\section{Spatial sampling and magnitude completeness}

For many cluster analyses, it is important that the candidate
galaxies in the cluster fields are evenly sampled.  
Radial variations in sampling will result in misrepresentation of
the cluster velocity dispersion and mass if there is a non-constant 
dispersion profile.  Azimuthal variations can result in overlooking
subclusters that are kinematically distinct. Splotchy sampling in
general can cause false positives in substructure tests that are 
sensitive to bimodality or asymmetry (Pinkney \etal 1996).  
Thus, one should be aware of the spatial uniformity of one's
cluster redshift survey.
Moreover, if one wishes to compare the occurrence
rate of substructure in one survey to another, it is desirable
to know the depth and area of coverage in both surveys.
We will address these issues as follows:
first, we describe how our lists of galaxy candidates were created,
second, we will estimate the completeness of these lists
(hereafter, {\it MX files}),
and third, we will evaluate how evenly we placed fibers on the 
candidate galaxies.

\subsection{Candidate galaxy selection}
The cluster search fields were roughly square with sides 
of 3\hub\ Mpc (Table \ref{spatial} gives dimensions) and
centered on the WAT.  A digitized image was obtained from the 
``Quick-V" survey (DSS).
Galaxies were selected by inspecting prints of the 
Palomar Sky Survey (PSS) with a binocular magnifier.
Objects were circled on a hardcopy of the DSS image 
and given priorities to be used in fiber placement.  
We included galaxies below the magnitude limit for reliable 
cross-correlation (estimated from early data). We also
included faint, compact objects which showed slight
``fuzz" or asymmetry.  
The poor candidates were given low priorities so 
that obvious galaxies would be observed first.
Coordinates were then measured to $\simlt$ 0\parcs5 
precision on the DSS images 
using the Guide-star Astrometric Support Package (GASP).
These coordinates were assembled into an {\it MX file} for
each cluster.
Finally, the MXPACKAGE tasks in 
IRAF\footnote{IRAF is distributed by the National Optical
Astronomy Observatories, which is operated by AURA under cooperative
agreement with the NSF.} 
are used to select targets for individual exposures.
We iterated on the priorities after observing runs to eliminate
the stars.

\subsection{Completeness of candidate galaxies and depth of spectra}

To address the completeness of our {\it MX files} 
we used new CCD data 
obtained for Abell 690 at the M-D-M 1.3-m on Jan 4, 1999.
We used a 2048$^{2}$ STIS CCD with 0.44 $''$/pix.   % Nellie
We obtained 1160 sec of total exposure in each of the V and R filters.
The dithered exposures were combined to make a 14\parcm4 x 15\parcm0
image (1.15 x 1.20 \hub Mpc).  The gibbous Moon made 
the images noisy and 
difficult to flatfield; our limiting magnitude was R=21.2 (2.5$\sigma$).
The FWHM of the PSF was about 1\parcs8.

The detection and classification of objects in these frames were done
using SExtractor (\cite{Bert96}) in double-image mode.  
SExtractor provides a ``stellarity
index" (hereafter, $SI$) which corresponds roughly to a confidence limit
(1=stellar, 0=non-stellar).
SExtractor could rule out
stellarity with 95\% confidence ($SI\leq0.05$) down to R=19.0. 
We used an adaptive aperture magnitude 
which included all but \app10\% of the light for galaxies.  
A photometric calibration using Landolt standard stars allowed us
to set the zeropoint.
A comparison to 11 field stars in the USNO\_A01 
(Monet et al. 1996) gave a zero point offset of 0.16 mag,  which is
less than the USNO plate-to-plate errors. We also estimated 
our scatter to be $<$ 0.13 mag.

The total number of {\it MX file} objects in the CCD 
field was 79.
SExtractor classified 50 of these as non-stellar, with $ SI\leq$0.2,
18 as ambiguous objects (0.2$\leq SI \leq 0.8$), and 11 as
stellar (0.8$\leq SI \leq 1.0$).
The numerous ambiguous and stellar objects mostly appear faint 
and/or compact on the PSS and were given low priorities.
Most of the $SI\geq$0.8 objects are stars, but
we measured a C1 galaxy velocity for one object with $SI=0.98$.
Five more examples of C1 galaxies were found with $0.2\leq SI \leq 0.8$.
Had our {\it MX} candidates been selected based on these $SI$ results,
several real galaxies would have been omitted.
This underscores the possibility that compact galaxies
are often overlooked in redshift surveys (Drinkwater et al. 1999).
SExtractor also erred in the opposite sense, classifying stars
as galaxies, particularly where the PSF distorted near the field edge. 
The 3 brightest of these were omitted from the following.

\placefigure{fig4}

The completeness of our A690 {\it MX file} is plotted in Fig \ref{fig4}.
The solid line is the fraction of CCD galaxies that are
included in the {\it MX file}.
There are 4 omissions in the R=16.75 and 17.25 bins.  Two of these
galaxies were in fact spotted on the PSS, but failed to be catalogued
because of mistakes in bookkeeping. 
One was excluded from MX because of its proximity to a bright star. 
One CCD object was overlooked.
The fainter, R=17.75 bin is 100\% complete.
In R=18.75, the completeness of our {\it MX file} is about 70\% ,
and it quickly drops thereafter.
Combining all bins with R$\leq$ 19.0, the A690 {\it MX file} is 82\% complete.
Thus, we find that our method of galaxy selection is
also subject to errors, but few of these are oversights.

We suspect that our other {\it MX files} were not equally deep.
Table \ref{spatial} shows the
number of candidate galaxies for each cluster, ``N$_{Cand.}$",
which is a measure of how thoroughly the PSS prints were 
searched for candidates. However, it must be 
modulated by the the cluster richness count (R), 
search field size (Field), field star density, and redshift (Table \ref{elg}).  
It appears that A1684 and A2220 have relatively small candidate
lists for their richness, field size, etc., and thus are likely 
to have lower completeness.  In contrast, A2462 has
a relatively large N$_{Cand.}$.  

If all {\it MX files} are far deeper than the limit of the detector,
the variations are inconsequential.
We have examined the limit of detection by the 2.3-m + MX using
the A690 CCD data.  We consider a ``detection" to be a
spectrum with a reliable (C1) redshift.
We do not expect a sharp magnitude limit because the detectability
should vary with observing conditions, the compactness and spectral type
of the galaxy, etc.
We find that, for 1 hour
exposures, about 50\% of objects with 18.0 $< R < $18.5 provided C1 
redshifts.  Below this, the number of C1 results declines rapidly.  This
suggests a rough limit of R\app 18.5. However, our probabilities are 
boosted in the case of A690 because most of its galaxies were 
observed repeatedly.  Thus, R\app 18.5 is the limit
under the best conditions.
For a single exposure in average conditions, the magnitude of 50\% 
probability is probably brighter than R$\sim$18.0 (Cf. R\app 17.25, Pi94).
Since over half of the {\it MX file} objects in the A690 CCD field were 
fainter than R=18, our {\it MX files} appear deeper than the MX limit.
Also, our target priorities cause the brightest galaxies to
be observed first.
Thus, any variations in the depth at which we surveyed the PSS prints 
should not seriously affect the depth of our detections. 

We have shown that the {\it MX files} are nearly complete
down to the detection limit of the instrument (R$\sim$18.5).  
But how many objects were actually {\em observed} 
to this depth, and how many yielded reliable redshifts?
The dotted line in Fig. \ref{fig4}
indicates the fraction of A690 objects with $SI\leq0.2$ 
for which a spectrum was obtained.
We have a spectrum for 68\% (34/50) of all $SI\leq0.2$ objects brighter 
than R=18.5,   
and a C1 redshift for 50\% .
Since this sample defined by SExtractor omits at least 6 galaxies 
and includes stars, let us take
the sample of 79 {\it MX file} objects contained within the CCD field
and remove objects with $SI\geq0.2$ and R$>$18.5.  
Then we have a spectrum for 95\% and a C1 redshift for
61\%.
Lacking a definitive list of galaxies in the field, we estimate
that about 60\% of the galaxies in A690 have reliable redshifts 
down to the MX limit (R=18.5).

The other WAT clusters may differ in completeness.
Table \ref{spatial} 
shows the {\it sampling percentage} (``\%"), or  
the percentage of {\it MX file} objects that were observed.  
For A690, the sampling percentage is 76.4\%, which ranks 3rd from highest. 
The 1st and 2nd ranked percentages are for A1684 and A2220, but since these
have a relatively small number of candidates (N$_{Cand.}$),
A690 is probably deeper.  
A690 had 19 exposures which should boost its completeness considerably.
The clusters with the {\em lowest} sampling percentages,
A1446 (42.4\%) and A2214 (52.1\%), are likely to have lower
completeness levels because of the low number of exposures. 
The cluster with the lowest percentage, A2462 (34.5\%), may
not be especially incomplete because it has a very
large catalog (N$_{Cand.}$=444) for its richness (0).

In summary, the completeness of our A690 {\it MX file} is
82\% down to R=19.0 and other clusters should scale roughly
by N$_{Cand.}$.
The sampling of our well-observed clusters (e.g., A690)
is very high (about 95\% to R=18.5) resulting in reliable
redshifts for about 60\%. Unfortunately, 
some of our clusters require more observations for this level of
completeness.  
For example, we are in danger of missing substructures
in A1446 and A2214 where the sampling drops below 40\% in 
subregions.  

\placetable{spatial}

\subsection{Spatial fairness of sampling}

We were concerned with obtaining uniform spatial coverage of our clusters.  
The priorities we assigned to the objects in our {\it MX files} were used
to control this.
MXPACKAGE would reduce the priorities to avoid repetitions
in consecutive exposures, and we manually reduced the
priorities from night to night to better sample all the galaxies.
We further influenced the spatial coverage 
by the choice of a center star for guiding.  This made different
regions of the field accessible to the probes.
Finally, the density of fiber probes as a function of radius was also  
adjustable by the user.  

We have evaluated how evenly we sampled the candidate galaxies
by dividing the cluster search fields into 6 subregions.  
These subregions are shown in Figure \ref{fig5} for A690
and A2634.  
The subregion sampling percentages for each cluster are
summarized in Table \ref{spatial}. 
Abell 1684, and 2214 appear to be undersampled in their 
centers compared to their outer regions. The difference between
these two counts is only significant at the 1.3 and 2.3 $\sigma$ 
level, respectively.  
These are fairly distant clusters with
less than 5 exposures, so their cores were not thoroughly probed.
Abell 2462 and 2634 
are {\em over}-sampled in the center compared to the outer regions
at the 4.9 and 3.1 $\sigma$ levels, respectively.
A2634 also shows a significantly deficient SW quadrant.  
Abell 160 is marginally over-sampled in the center (2.8 $\sigma$).
The other clusters (including A690 in Fig. \ref{fig5})
are reasonably evenly sampled.  
Thus, we find 3$\sigma$ non-uniformities in 2 cases, A2462
and A2634, and 4 other less significant cases: A160,
A1684, A2214, and A2220.
Even the less severe non-uniformities could influence 
substructure analysis and velocity dispersion measurements.
We will keep this in mind for our analysis in paper II.  

\placefigure{fig5}

% section 5
\section{Summary }

We have presented new redshifts and positions for galaxies
in 9 WAT clusters. Nineteen of our galaxies had published redshifts
which are in reasonable agreement with our redshifts (i.e., $\sigma \simlt 
150$).
A search for emission line galaxies helped contribute cluster redshifts
and improve the accuracy of our catalog.
Our candidate galaxies for observation were chosen by visual inspection
of the PSS. CCD data for Abell 690 indicate that its candidate list 
({\it MX file}) is 82\% complete down to R$\sim$19.0. Also,
95\% of A690 targets are observed, while about 60\% of them have reliable
redshifts down to the MX detection limit (R$\sim$18.5).
The fraction of candidates that are {\em observed} varies from 
cluster to cluster, and, for 2 clusters, varies strongly between
subregions in the cluster.
Nevertheless, the dataset is well-suited for measuring WAT peculiar velocities,
and can also be used for substructure analyses with attention
to the cases with uneven sampling.
Paper II will use these data to look for links between the WAT radio morphology 
and the cluster's dynamical state.  

\acknowledgements

We would like to thank George Rhee for assistance on the initial
observing runs,
William Oegerle for providing digitized, ``Quick V" survey images
which were essential for galaxy astrometry, the Steward Observatory 
TAC, and Frazer Owen, who kindly made available a listing of sources from
the 20 cm VLA Northern Abell Cluster Survey prior to publication.
Support was provided for this research by 
NSF grants AST-9317596, AST-9896039, and NASA grant
NAGW-3152 to J. O. B., and by HST grant
GO-07388.01 and LTSA grant NAG5-8238 to D. Richstone.
Data were obtained (in part) using the 1.3 m McGraw-Hill Telescope
of the M-D-M Observatory.
This research has made use of the NASA/IPAC Extragalactic Database
(NED), which is operated by the Jet Propulsion Laboratory, 
California Institute of Technology, under contract with the
National Aeronautics and Space Administration.

%%%%%% REFERENCES HERE %%%%%%%%%%%%%%%%%%%%%%%%%%%%%%%%%%%%%%%%%%%

%%%%%% REFERENCES ABOVE %%%%%%%%%%%%%%%%%%%%%%%%%%%%%%%%%%%%%%%%%%%

%%%%%% FIGURE CAPTIONS HERE %%%%%%%%%%%%%%%%%%%%%%%%%%%%%%%%%%%%%%%%%%%
\clearpage

\begin{center}
{\bf List of Figures}
\end{center}

\figcaption[specall.ps]{Spectra from the MX multifiber spectrometer.
The night sky spectrum has been subtracted, but the
residuals of prominent lines (marked ``ns") may be visible. 
a) The WAT in Abell 1684. Note the interference of night sky with
Mg b. 
b) The WAT in Abell 1446.  
c) A convincing emission line spectrum from A98. 
d) A galaxy in A98 showing prominant Balmer absorption left of the
4000\AA\ break and slight OII emission indicative of post- or ongoing-
star formation.  
\label{fig1} }

\figcaption[histall1.ps]{Histograms showing all C1 velocities (see text)
measured in the fields of Abell 98, 160, 690, 1446, and 1684. The
number of C1 velocities is shown. \label{fig2}}

\figcaption[histall2.ps]{Histograms showing all C1 velocities 
measured in the fields of Cl1313+073, Abell 2214, 2220, and 2462. 
The number of C1 velocities is shown. \label{fig3}}

\figcaption[complete.ps]{Plot of completeness vs. SExtractor R magnitude
within a 14\parcm4X15$'$ field of Abell 690. Use the left axis for the
two completeness curves and the right axis for the triangles and squares.
The solid curve is the fraction of SExtractor galaxies ($SI\leq$ 
0.2) which are in the {\it MX file} (see text). 
The dotted curve is the fraction of the SExtractor galaxies which were 
actually observed by MX.  The boxes are the number of SExtractor galaxies 
in each 0.5-mag bin, while the triangles are the number of those SExtractor 
galaxies which are also in the {\it MX file}.
\label{fig4}}

\figcaption[perc.ps]{
Two plots showing the uniformity of spatial sampling for Abell 690
and Abell 2634.
The filled circles are candidate galaxies with a spectrum (i.e.,
``sampled") and the open circles are candidates without a spectrum.
The percentage of candidates observed is labelled inside six sub-regions: 
four equal-sized quadrants, a region inside of the circle (D = 1/2 
field height), and a region outside of the circle.  These sampling
percentages are given for all of the clusters in Table \ref{spatial}.
Note that Abell 2634 is undersampled in the lower right quadrant 
(Q4) and oversampled in the center.  
\label{fig5}
}

%%%%%% TABLE CAPTIONS HERE %%%%%%%%%%%%%%%%%%%%%%%%%%%%%%%%%%%%%%%%%%%
\clearpage

\begin{center}
{\bf List of Tables}
\end{center}

\dummytable
\label{tabwatrad} 
\noindent
Table 1: Radio Properties of WATs.

% Table 2
%  MX Observing Log   (See mxlog.tex in TABLES)
\dummytable
\label{mxlog} 
\noindent
Table 2: Log of MX Observing Runs.

% Table 3
% {Big Velocity and position table}
% \include{TABLES/bigtab}
\dummytable
\label{bigtab}
\noindent
Table 3: Positions and Velocities of WAT cluster galaxies.

% Table 4
% Velocity means and stddeviations
\dummytable
\label{means}
\noindent
Table 4: Traditional Velocity data for WAT clusters.

% Table 5
% Velocity comparison table
\dummytable
\label{vcomp}
\noindent
Table 5: Comparison with Published Velocities.

% Table 6
% {Emission Line Galaxy Counts}
% \include{TABLES/bigtab}
\dummytable
\label{elg}
\noindent
Table 6: Emission Line Galaxy Counts.

% Table 7: results of spatial sampling investigation
\dummytable
\label{spatial}
\noindent
Table 7: Spatial Fairness of MX Observations.

\end{document}